%% file: main.tex
\documentclass[conference]{IEEEtran}
\IEEEoverridecommandlockouts
\usepackage[T1]{fontenc} 
\usepackage{amsmath}
\usepackage{dsfont}
\usepackage[cmintegrals]{newtxmath}
\usepackage{bm} 
\usepackage{url}
\usepackage{mdwtab}
\usepackage{xcolor}
\usepackage[hidelinks]{hyperref}
\usepackage[capitalise]{cleveref} 
\makeatletter
\let\MYcaption\@makecaption
\makeatother
\usepackage[font=footnotesize]{subcaption}

\makeatletter
\let\@makecaption\MYcaption
\makeatother

\usepackage[acronym,nopostdot,style=index,nonumberlist,toc]{glossaries}
\usepackage{graphicx}

\usepackage{svg}
\usepackage{mathtools}
\usepackage{makecell}

\usepackage{breqn}   

\usepackage{booktabs}
\usepackage{float}
\usepackage{tabularx}
\input{acronyms}
\input{symbols}


\renewcommand\baselinestretch{.97}

\newcommand{\nonl}{\renewcommand{\nl}{\let\nl\oldnl}}
\let\oldnl\nl
\begin{document}

\title{Joint Channel and Semantic-aware Grouping for Effective Collaborative Edge Inference}

\author{
  \IEEEauthorblockN{
    Mateus P. Mota,
    Mattia Merluzzi,
    Emilio Calvanese Strinati,
    }
  \IEEEauthorblockA{CEA-Leti, Univ. Grenoble Alpes, F-38000 Grenoble, France\\
    Email: \{mateus.pontesmota, mattia.merluzzi, emilio.calvanese-strinati \}@cea.fr
   }
   \thanks{This work has been supported by the SNS JU project 6G-GOALS under the EU’s Horizon program Grant Agreement No. 101139232}
}
\maketitle

\begin{abstract}
We focus on collaborative edge inference over wireless, which enables multiple devices to cooperate to improve inference performance in the presence of corrupted data.
Exploiting a key-query mechanism for selective information exchange (or, group formation for collaboration), we recall the effect of wireless channel impairments in feature communication.
We argue and show that a disjoint approach, which only considers either the semantic relevance or channel state between devices, performs poorly, especially in harsh propagation conditions.
Based on these findings, we propose a joint approach that takes into account semantic information relevance and channel states when grouping devices for collaboration, by making the general attention weights dependent of the channel information.
Numerical simulations show the superiority of the joint approach against local inference on corrupted data, as well as compared to collaborative inference with disjoint decisions that either consider application or physical layer parameters when forming groups. 

%

\end{abstract}

\begin{IEEEkeywords}
Collaborative inference, Semantic and goal-oriented communications, edge artificial intelligence.
\end{IEEEkeywords}

\glsresetall

\glsresetall

\section{Introduction}
\label{sec:Intro}

Edge intelligence (or, edge \gls{ai}) systems are increasingly composed of lightweight, resource-constrained devices equipped with pre-trained \gls{ml}/\gls{ai} models capable of performing local inference on data captured from their surroundings.
These edge nodes, often lightweight and bandwidth-constrained, are designed to operate autonomously, making decisions in real time without needing to offload data to centralized servers.
%
%
Nonetheless, its reliability hinges on several factors: the quality of sensory data, the robustness of the deployed models, and the conditions of the wireless channel, especially when communication is involved to enable collaboration horizontally and vertically.

In isolation, a single device’s inference capability can deteriorate under such uncertain conditions.
It is in the scenario that distributed collaboration can provide a promising avenue leveraging the network of edge systems.
When devices share intermediate representations or contextual cues, they can collectively compensate for gaps in individual perception.
However, this cooperative process must contend with constraints across multiple layers—ranging from fluctuating wireless link quality to the heterogeneity of device roles and data relevance.
Effective collaboration, therefore, hinges not just on what is being shared, but with whom and under what conditions.

%


%
While this strategy offers notable benefits, including fault tolerance and reduced decision latency, it introduces a complex design space. 
Challenges span multiple layers: selecting which devices to involve, identifying what information to share, and managing communication cost—all while respecting device heterogeneity and varying link conditions.
This work approaches collaborative inference from a cross-layer perspective, where both application-level relevance and physical-layer channel quality guide information exchange.
Our methodology builds on the principles of semantic and goal-oriented communications \cite{Strinati24}, aiming to align data sharing with task-specific utility—an emerging requirement in 6G and AI-native wireless systems.



\vspace{4pt}

\noindent\textbf{\textit{Related Work}}
A key obstacle in collaborative inference lies in determining both what features to share and which peers to collaborate with.
Our work is inspired by approaches in collaborative perception \cite{han2023collaborative}, though we assume the inference model is fixed and pre-trained, rather than jointly learned during deployment.
For instance, \cite{liu2020when2com} proposes a query-key mechanism to learn communication graphs dynamically based on feature relevance.
Similarly, semantic-based data sourcing in \cite{huang2023semdas} and its extension to random access protocols in \cite{kalor2023random} exploit query-key matching to guide information exchange.

However, these methods grouping decisions rely solely on semantic matching, without factoring in the potential degradation introduced by imperfect communication links. 
While in \cite{Mateus25Eusipco} we study the effect of communication impairments and collaboration layer on the overall performance, we do not incorporate the link quality into the semantic method.
%

\vspace{4pt}

\noindent\textbf{\textit{Contributions}}
This paper considers a scenario where inference is executed at the edge, in line with the collaborative communication framework of \cite{liu2020when2com}, but explicitly accounts for wireless transmission impairments.
We extend the semantic matching procedure by incorporating channel state information into the learning pipeline, enabling the formation of communication graphs that are robust to both semantic and physical constraints.
Specifically, our contributions include:
\begin{itemize}
\item A novel joint grouping mechanism that leverages both semantic relevance and wireless channel quality;
\item An in-depth evaluation of the trade-off between communication cost and inference accuracy under realistic channel impairments.
\end{itemize}

%
%



This remainder of the paper is organized as follows:
\Cref{sec:system-model} describes the system model, \Cref{sec:solution} introduces the framework for communication grouping based on semantic matching.
Finally, \Cref{sec:results} details the experiments performed and discuss their results.
The work is then concluded in \Cref{sec:conclusion}.


\section{System Model}
\label{sec:system-model}

\begin{figure}[t]
      \centering

      \includegraphics[width=0.96\columnwidth]{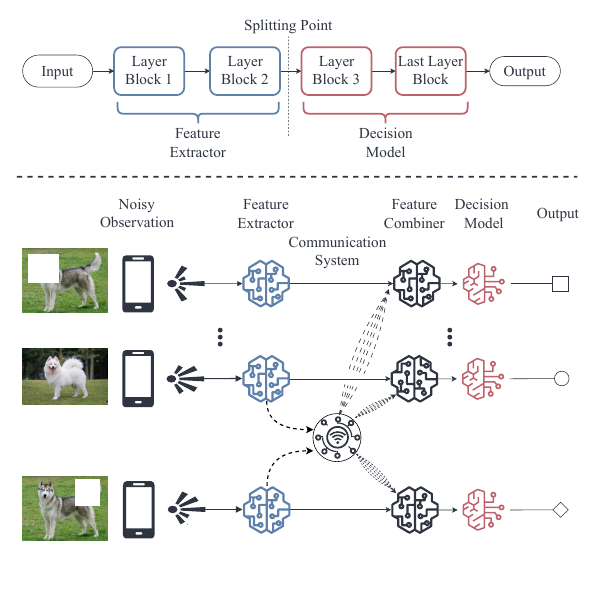}
      \caption{
            System model for the proposed collaborative inference problem: devices collect incomplete/corrupted data.
      }
      \label{fig:system-model}
\end{figure}

We use the system model of \cite{Mateus25Eusipco}, but we recall it for completeness.
The system is composed by \gls{not:nUE} devices empowered with \gls{ai} capabilities, in this case inference models.
Each device $i$ performs inference using a pre-trained model $\gls{not:full-model}$ on input $x_i$, e.g,. an image collected through a camera or other modality data.
The model is split into two parts, a feature extractor $\gls{not:encoder-model}$ and a decision model $\gls{not:decoder-model}$, as in \cref{fig:system-model}, such that:
\begin{dmath}
      \gls{not:full-model} (x) = \left( \gls{not:decoder-model} \circ \gls{not:encoder-model} \right) \, (x)
\end{dmath}.

We consider a scenario where the \gls{not:nUE} devices are organized into $\gls{not:nGroups}$ groups.
Devices belonging to the same group are assumed to observe identical input data.
That is, if devices $i$ and $j$ are both members of group $g$, then $x_i = x_j = x_g$.
%
For generality, we assume that the group assignment is random, and importantly, each device lacks knowledge of which other devices belong to its group.
The main objective is for devices to autonomously discover their group peers to improve local inference performance, with low overhead for the network, i.e. without the need of directly sharing data.


Each device observes either a noisy version $\hat{x}_i = M^i(x_g)$ of the true group data $x_g$ with probability $\gls{not:p-patch}$, or the full observation $x_g$ with probability $1 - \gls{not:p-patch}$.
Devices then transmit the output of their local encoder, $\gls{not:encoder-model}(\hat{x}_i)$, to peers via a generic communication system $\gls{not:comm-system}$.
%
%
%
This communication channel is treated abstractly—it could represent a direct wireless link, an edge-bound transmission, or any form of data transformation and transfer.
Due to potential imperfections in the channel, the data received by other devices may be corrupted or degraded compared to the original transmission.
%
%

For a device $i$, the information shared by other devices within the same group plays a crucial role in enhancing its local inference, particularly when its own observation is noisy or degraded.
However, this requires discovering the suitable devices for collaboration in terms of semantic relevance and good mutual channel conditions.
%
%
When device $j$ sends a general message $o_j$, the corresponding message received by device $i$ via $C$ is represented as:
%
\begin{dmath}\label{eq:rx_message}
    y_c^{ij} = \gls{not:comm-system} \left\{  \gls{not:observation}_i \right\}
\end{dmath},
where we use the notation $\{\cdot\}$ to denote a system, rather than a function.


The devices need to aggregate the information received by their created group.
This is performed by the feature combiner, $\gls{not:combiner-model}^i$ at device $i$.
Denoting by $\mathbf{y}_c^i$ the aggregated information received by device $i$ from the other devices in the group, the output of the feature combiner is
\begin{dmath}\label{eq:feat_comb}
    y^i_g = \gls{not:combiner-model} \left( \gls{not:encoder-model} ( \hat{x}_i ) , \mathbf{y}_c^i \right)
\end{dmath}.

Finally, the combined information is fed to the decision model to provide the inference result at device $i$:
\begin{dmath}
    y_d^i = \gls{not:decoder-model} (y^i_g)
\end{dmath}.
This whole procedure is illustrated in \cref{fig:system-model}.

The ultimate objective in this scenario is to enable collaborative inference, enhancing overall accuracy through the exchange of intermediate information via wireless communication.
To achieve this, devices can utilize the communication system not only to transmit their feature representations, but also to discover other devices that possess information pertinent to improving their own inference capabilities.
This process requires appropriately weighting each device's contribution within the feature combiner.

%
%

\subsection{Communication System}

As in \cite{Mateus25Eusipco}, we consider sidelink communication, which can occur in different ways:
\begin{itemize}
      \item \textbf{Unicast:} One-to-one communication. Data is transmitted to a single device.
      \item \textbf{Multicast:} One-to-many communication. Data is transmitted to dedicated set of devices in the area.
      \item \textbf{Broadcast:} One-to-all communication. Data is transmitted to all devices in the broadcast area.
\end{itemize}
Unicast and multicast communication need network connection, since they use the uplink to request communication, i.e. request to join a group.
\textit{Semantic queries} are exchanged via multicast transmission, while unicast transmission is used to exchange intermediate observation.

%
Differently from \cite{Mateus25Eusipco}, the communication system $C$ is a modeled as an \gls{awgn} channel, where ratio between the signal and the noise power is the \gls{snr}.
The \gls{snr} of the semantic query transmission from device $i$ to $j$ is denoted by $\gamma^{q}_{ij}$ while the \gls{snr} of the respective data transmission from $j$ to $i$ is given by $\gamma^{d}_{ji}$
We assume that the devices exchange pilots in order to have a sidelink channel estimation, and that the latter is perfect.
For the sake of generality, we assume that semantic query and the data transmission channels are different, so that $\gamma^{d}_{ij} \neq \gamma^{q}_{ij}$. This covers mobility scenarios.

%
%
%

\subsection{Key Performance Indicators}
%
%
%
Since we consider a classification task, we consider accuracy as KPI at the application layers, i.e., the goal of communication. This is typical of the semantic and goal-oriented approach, which does not solely focus on communication-related KPIs.
%
%
%
%
%
%
\begin{table*}[t]
      \begin{minipage}[t]{\columnwidth}
        \caption{Simulation Parameters}
        \centering
        \label{tab:sim-params}
            \begingroup
            \setlength{\tabcolsep}{12pt} 
            \begin{tabularx}{0.9\columnwidth}{l c c}
            \toprule
            \textbf{Parameter}     & \textbf{Symbol}  & \textbf{Value}           \\
            \midrule
            Number of devices      & \gls{not:nUE}           &  $16$                    \\
            Number of groups       & \gls{not:nGroups}       &  $4$                     \\
            Data size (\# of floating-points) &                & $4000$ \\
            Key size               & K                       &  $1024$                  \\
            Patch scale            &                         &  $0.4$                   \\
            Probability of partial observation  &   \gls{not:p-patch} &  $0.8$        \\
            Maximum SNR            & $\gamma_{\mathrm{max}}$ &  $10$dB                  \\
            Minimum SNR            & $\gamma_{\mathrm{min}}$ &  $-10$dB                  \\
            \bottomrule
            \end{tabularx}
            \endgroup
      \end{minipage}\hfill 
      \begin{minipage}[t]{\columnwidth}
        \caption{Training Parameters}
        \centering
        \label{tab:train-params}
            \begingroup
            \setlength{\tabcolsep}{10pt} 
            \begin{tabularx}{0.9\columnwidth}{l c}
            \toprule
            \textbf{Parameter}       & \textbf{Value} 	            \\
            \midrule
            Batch size          & $64$                             \\
            Number hidden layers      & $ 2 $      \\
            \# of neurons per hidden layer in \gls{not:generator-key} and \gls{not:generator-query}     & $ \left[ 256, 128 \right] $      \\
            Activation function of hidden neurons    & ReLU      \\
            Optimizer algorithm     & Rectified Adam    \\
            \# of epochs         & $60$                     \\
            Learning rate        & $ 10^{-5}$                     \\
      
          \bottomrule
          \end{tabularx}
          \endgroup
        \end{minipage}
      \end{table*}

\begin{figure*}[!tb]
      \centering
    \begin{subfigure}[t]{\columnwidth}
        \centering
        \includegraphics[width=\textwidth]{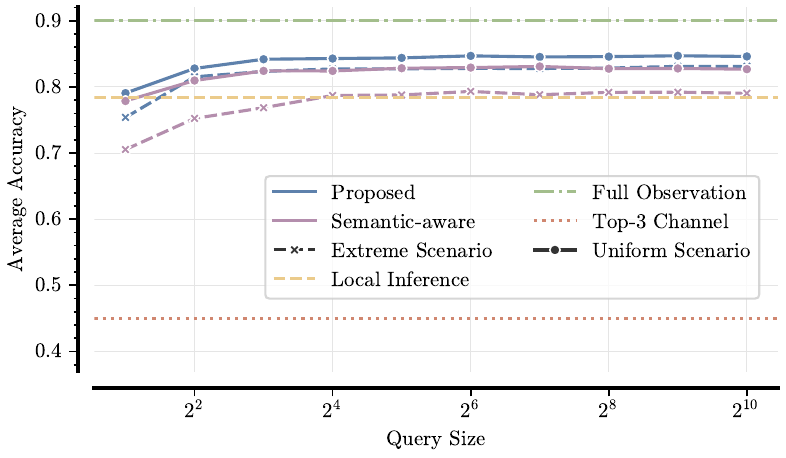}
        \caption{Comparison of performance when intermediate data transmission is affected by channel errors.}
        \label{fig:query-data}
        \end{subfigure}%
    \hfill
    \begin{subfigure}[t]{\columnwidth}
        \centering
        \includegraphics[width=\textwidth]{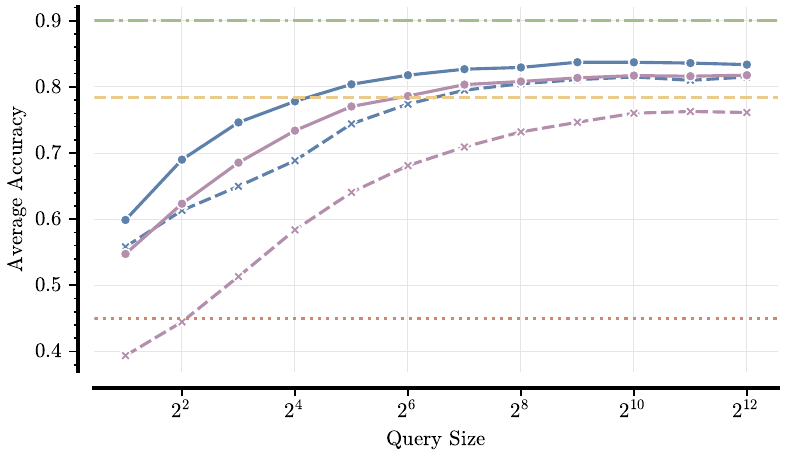}
        \caption{Comparison of performance when intermediate data and query transmissions are affected by channel errors.}
        \label{fig:query-full}
    \end{subfigure}
    \hfill
    \caption{Comparison of the different methods and scenarios in terms of accuracy for different query sizes.}
    \label{fig:results-query}
\end{figure*}

%
%
However, the cost in terms of communication and computation is also a fundamental metric to be taken into account. Therefore, as a second KPI, we consider the average resource usage in terms of sidelink connections.
%
%
The latter is computed as the average number of sidelink transmissions resulting from the optimized device grouping (i.e., the communication graph).
%

\section{Link-aware semantic matching-based grouping}
\label{sec:solution}


Given the above task, this work leverages an attention-based mechanism similar to \cite{liu2020when2com} to identify: i) whether a device needs extra information for inference due to corrupted local data, and ii) the set of devices to collaborate with, in case of bad quality local data.
%
%
Adapting the framework to our system, device $i$ compresses its observation, obtaining an intermediate representation $\gls{not:observation}_i = \gls{not:encoder-model} \, (\hat{x}_i)$.
Then, it generates: \textit{i)} a low-dimensional query vector $\gls{not:vec-query}_i$ and \textit{ii)} a key vector $\gls{not:vec-key}_i$:
\begin{gather}
      \gls{not:vec-query}_i = \gls{not:generator-query} (o_i ; \theta_q ) \, \text{,} 
      \\
      \gls{not:vec-key}_i = \gls{not:generator-key} (o_i ; \theta_k) \, \text{,}
\end{gather}  
\noindent where $\gls{not:generator-key}$ and $\gls{not:generator-query}$ are two neural networks parametrized by $\theta_k$ and $\theta_q$, respectively.
The query is transmitted to all other devices, while the key is kept local.
The query received by device $j$ from device $i$ is $\hat{ \gls{not:vec-query}}^j_i=C\{\mu_i\}$.

%
Every device receives the queries of all others (multicast), and uses its key to compute a matching score through scaled general attention \cite{luong2015effective}.
This matching score represents the relevance of a device information to another device, so that they can exchange data, or, their intermediate representation (unicast) and weight the contribution accordingly.
Differently from previous works, we propose to include the wireless link information in this step, so that the matching score also weights its effect.
We denote by $m_{i j}$ the matching score for device $j$ receiving query from device $i$, which reads as 
\begin{equation}
\label{eq:general-attention}
      m_{ij} = \frac{{\gls{not:vec-key}_i}^{\intercal} \gls{not:attention-weights}^j_i \hat{\gls{not:vec-query}}^j_i}{\sqrt{K}} \, \text{,}
\end{equation}
\noindent where originally $\gls{not:attention-weights} \in \mathbb{R}^{Q \times K}$ is a trainable parameter without any dependencies to match the query size, $Q$, and the key size, $K$.
However, we propose to make $\gls{not:attention-weights}$ dependant of link information as:
\begin{equation}
    \label{eq:weight-generator}
    \gls{not:attention-weights}^j_i = \gls{not:generator-weight} (\gamma^{d}_{ij}, \gamma^{q}_{ji})    \, \text{,}
\end{equation}
\noindent where $\gamma^{d}_{ji}$ and $\gamma^{q}_{ij}$ represent, respectively, the \gls{snr} of the data and query transmission from device $j$ to $i$.


%
All the matching scores are used to construct a matching matrix $\mathbf{M}$ by using a row-wise softmax, with elements $\bar{m}_{ij}$.
The latter is used to construct the communications graph, as its values $\bar{m}_{i j}$ represent how relevant the information of device $i$ is for device $j$, recalling that this relevance pertains to the semantics and the wireless channel conditions.
Once the groups are created, the devices share the actual data (or, their intermediate representation), which might have high dimensionality compared to the queries.
To avoid high communication overhead, $\mathbf{M}$ can be pruned with threshold $\rho$, i.e., $\bar{m}_{ij}^\rho=\bar{m}_{ij}\cdot\mathbf{1}\{\bar{m}_{ij}\geq\rho\}$, where $\mathbf{1}\{\cdot\}$ denotes the indicator function.
$\mathbf{M}$ is also used to combine features (cf. \eqref{eq:feat_comb}) according to the following weighted average:
\begin{equation}
      y^i_g = \sum_{j=1}^{\gls{not:nUE}} \bar{m}_{i j}^\rho y_c^{ij} \, \text{.}
\end{equation}
where $y_c^{ij}$ is the received version of the intermediate data, as per the definition in \eqref{eq:rx_message}.

We can summarize the procedure as follows:
\begin{enumerate}
    \item All devices generate a key $\gls{not:vec-key}_i$ and a query $\gls{not:vec-query}_i$ based on their intermediate representation. 
    \item A device $i$ transmits its query to all other devices.
    \item When in possession of the received query $\hat{ \gls{not:vec-query}}^j_i$, a device $j$ calculates the matching score $m_{ij}$ taking into consideration the wireless link information.
    \item Device $j$ transmits its intermediate representation to $i$ if $\bar{ m}_{ij} \geq \rho$
    \item Device $i$ aggregates the received data according to the weights $\bar{m}_{ij}$.
\end{enumerate}

In this work, we consider image classification as application. As such, training is performed by computing the cross-entropy loss between the true label and the predicted label, $y_d^i = \gls{not:decoder-model} (y^i_g)$.
%
It is important to highlight that, differently from \cite{liu2020when2com}, only the query generator $\gls{not:generator-query}$, the key generator $\gls{not:generator-key}$ and the attention weights $\gls{not:generator-weight}$ are learned.
%
As such, the encoder model $\gls{not:encoder-model}$ and the decoder model $\gls{not:decoder-model}$ are assumed to be pre-trained and their parameters frozen, while only the modules needed for the communication need to be trained.
As a consequence, the pre-trained encoder and decoder models are shared across all devices.
Note that decentralized training is also possible, with the result of different models for each device. However, this increases the computational cost.
%
%
\section{Numerical results}
\label{sec:results}

\subsection{Simulation setting and parameters}

Image classification is performed on the Imagenette dataset \cite{Howard_Imagenette_2019}.
The pre-trained model is the MobileNetV3-Small \cite{howard2019searching}, initialized with its default weights from training on the ImageNet dataset and then fine-tuned to the Imagenette dataset.
The partial data observability (i.e., the local data corruption) is modeled by applying a white patch in a random position of the image, with the ratio between the white patch size and the image size being $0.4$.
In other words, $40$\% of the image is locally missing at the device, if the latter belongs to the set of devices with corrupted data.
%

We consider two wireless condition scenarios:
\begin{itemize}
    \item \textbf{Uniform scenario:} The \glspl{snr} are sampled from a uniform distribution in the interval $\left[ \gamma_{\mathrm{min}}, \gamma_{\mathrm{max}} \right]$.
    \item \textbf{Extreme scenario:} The \glspl{snr} are sampled from a discrete uniform distribution with values $\left\{ \gamma_{\mathrm{min}}, \gamma_{\mathrm{max}} \right\} $.
\end{itemize}
%

%
%

\subsection{Baseline Solutions}

The semantic grouping solution, with and without link information (\textit{Proposed} and \textit{Semantic-aware} in the legend) is compared with three other benchmark solutions:
\begin{itemize}
      \item \textbf{Local inference:} only the local observation is used for inference, possibly on corrupted data. This represents the non-collaborative case.
      \item \textbf{Full observation:} Inference is performed with the true observation instead of the partial one. This represents a performance upper bound.
      \item \textbf{Top-3 Channel:} Each device selects the three devices with best sidelink channels and it receives their information, which is averaged with the same weights. Not based on the semantic data relevance.
\end{itemize}
%



\subsection{Channel effect on query and data transmission}
%
First, we analyze the effect of considering the link information in the method, comparing the semantic grouping solution trained for different query sizes with the other baselines in terms of accuracy.
%
%
These results are shown in \cref{fig:results-query}, with the results in \cref{fig:query-data} assuming that the query transmission is reliable and only the intermediate data transmission is affected by channel noise, whereas in \cref{fig:query-full} the query transmission is also subject to noise.
%

\begin{figure*}[!tb]
      \centering
      \begin{subfigure}[t]{\columnwidth}
            \centering
            \includegraphics[width=0.95\textwidth]{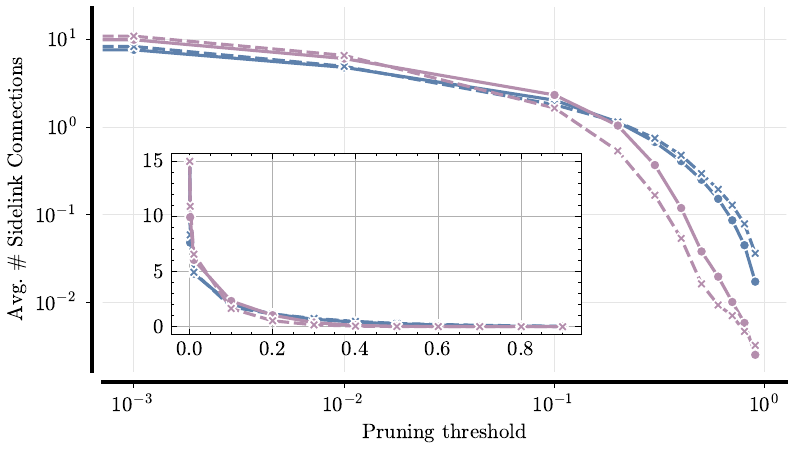}
        \caption{Effect on the average number of sidelink connections per device}
        \label{fig:conn-thr}
        \end{subfigure}%
    \begin{subfigure}[t]{\columnwidth}
          \centering
          \includegraphics[width=0.95\textwidth]{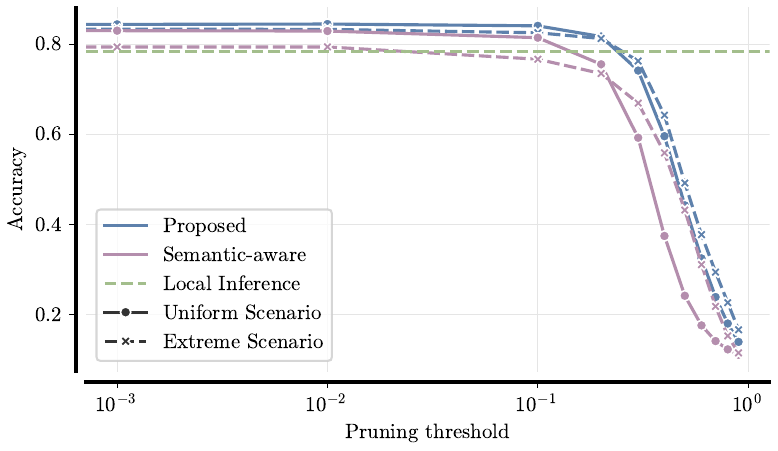}
          \caption{Effect on the accuracy}
          \label{fig:acc-thr}
    \end{subfigure}
    \caption{Studying the effect of the communication pruning trheshold}
    \label{fig:results-thr}
  \end{figure*}

%
From \cref{fig:query-data}, we notice how adding the wireless link information improves performance, for both scenarios, with the collaboration improving performance when compared to standalone inference on local corrupted data or with considering only the link quality to select sources.
It is also important to highlight that the query is of small size (in bits) with respect to the data, with the performance plateauing with a query size of $64$ for the hardest case (extreme scenario without channel information).
However, \cref{fig:query-data} shows that the effect of the channel on the query transmission makes the task much harder, as it needs a larger query to surpass the performance of the local-only solution.
It also shows the importance of considering link information, as the link-unaware solution is unable to outperform local inference in the extreme scenario.

These results suggest that the proposed solution is able to effectively weight both the data relevance and the channel conditions, improving the semantic matching method.
They also suggest that the query needs a reliable transmission scheme, compared to the data (or intermediate representation) itself.
We can conclude that semantic representation helps communication robustness, but device grouping needs reliable query exchange to achieve acceptable performance.
However, it should be noted that, \textbf{given the small size of the query, increased communication reliability effort does not impact system cost as data transmission itself}.
We also note that grouping based only on the channel does not perform well since it fails to filter the data based on its relevance to a device.

\subsection{Communication pruning effect}
%
We now analyze the effect of the communication pruning, which reduces the communication by only transmitting information if the matching score is above a certain threshold.
These results are shown in \cref{fig:results-thr}, assuming a scenario with noise affects only the data transmission, as in \cref{fig:query-data}, and with a query size of $1024$.
In \cref{fig:conn-thr}, we plot the average number of sidelink connections per inference task, as a function of the pruning threshold $\rho$.
Whereas \cref{fig:acc-thr} shows the corresponding accuracy, also as a function of $\rho$.

\cref{fig:conn-thr} shows that the lowest (positive) threshold already provides significant communication reduction.
This is thanks to the fact that some of the elements of the matching matrix are very close or equal to zero.
Naturally, this does not reduce accuracy, as shown in \cref{fig:acc-thr}, because only lowly weighted collaborations are pruned.
\cref{fig:conn-thr} also shows that pruning further reduces communication when the channel information is included (our semantic and link-aware approach).
%
This is shown by the communication reduction for low pruning thresholds close to zero.
It implies that when the link quality information is not available, the solution relies on more sources to counter the unknown channel conditions, whereas the proposed improvement allows to filter the sources by link quality, thus having more lower valued matching scores.
The extreme scenario needs an even stronger source selection, as it experiences less connections for the same threshold.
This suggests that the proposed solution allows a better semantic matching by considering the channel conditions.

Comparing \cref{fig:conn-thr} and \cref{fig:acc-thr}, we can conclude that it is possible to reduce communication effectively without affecting accuracy.
However, if communication is heavily reduced, the degradation in performance can even overcome local performance with corrupted data, making collaboration useless.

\section{Conclusions and Perspectives}
\label{sec:conclusion}


This work presents a step toward robust and efficient collaborative intelligence at the wireless edge by investigating how communication-aware design can enhance inference reliability in distributed AI systems.
Instead of handling semantic relevance and channel conditions separately, we unify them through a joint decision-making framework that adapts dynamically to both the task relevance and the network environment, by adding the link information to the general attention weights.
%


Our results show that collaborative grouping strategies informed by both data relevance and link quality outperform traditional disjoint schemes, which often neglect the interplay between inference goals and physical-layer constraints.
This synergy proves especially beneficial under harsh wireless conditions, where smart source selection and communication pruning can sustain high inference accuracy with minimal overhead.
%
The proposed joint solution also reduces further the number of sidelink connections when compared with the semantic-aware solution, since it considers the link conditions when matching devices.
%
%

Looking ahead, future developments could explore query-conditioned feature encoding, enabling devices to tailor their outgoing messages based on the specific intent of incoming queries. 
Such query-aware compression could substantially reduce the bandwidth footprint of collaborative inference without compromising performance.


      
\bibliographystyle{IEEEtran}
\bibliography{IEEEabrv, ref.bib}

\end{document}

%% file: acronyms.tex
\newacronym{mmwave}{mmWave}{millimeter wave}
\newacronym{ai}{AI}{artificial intelligence}
\newacronym{xai}{XAI}{explainable artificial intelligence}
\newacronym{embb}{eMBB}{enhanced mobile broadband}
\newacronym{urllc}{uRLLC}{ultra-reliable and low-latency communications}
\newacronym{mmtc}{mMTC}{massive machine-type communications}
\newacronym{nr}{NR}{new radio}
\newacronym{5g}{5G}{fifth generation}
\newacronym{3gpp}{3GPP}{Third Generation Partnership Project}
\newacronym{bs}{BS}{base station}
\newacronym{aoi}{AoI}{age of information}
\newacronym{jsrl}{JSRL}{jump-start reinforcement learning}
\newacronym{awgn}{AWGN}{additive white Gaussian noise}
\newacronym{snr}{SNR}{signal-to-noise ratio}
\newacronym{ue}{UE}{user equipment}
\newacronym{leo}{LEO}{low-altitude earth orbit}
\newacronym{ee}{EE}{energy efficiency}
\newacronym{dci}{DCI}{downlink control information}
\newacronym{uci}{UCI}{uplink control information}
\newacronym{usc}{USC}{uplink slotted shared channel}
\newacronym{l2c}{L2C}{learning to communicate}
\newacronym{ir}{IR}{immediate reply}

\newacronym{dcm}{DCM}{downlink control message}
\newacronym{ucm}{UCM}{uplink control message}
\newacronym{pdcch}{PDCCH}{physical downlink control channel}
\newacronym{pdsch}{PDSCH}{physical downlink shared channel}
\newacronym{pucch}{PUCCH}{physical uplink control channel}
\newacronym{pusch}{PUSCH}{physical uplink shared channel}
\newacronym{rrm}{RRM}{radio resource management}
\newacronym{dlsch}{DL-SCH}{downlink shared channel}
\newacronym{bch}{BCH}{broadcast channel}
\newacronym{pch}{PCH}{paging channel}
\newacronym{ul}{UL}{uplink}
\newacronym{kpi}{KPI}{key performance indicator}
\newacronym{dl}{DL}{downlink}
\newacronym{ulsch}{UL-SCH}{uplink shared channel}
\newacronym{rach}{RACH}{random-access channel}
\newacronym{mac}{MAC}{medium access control}
\newacronym{tdma}{TDMA}{time division multiple access}
\newacronym{tti}{TTI}{transmission time interval}
\newacronym{phy}{PHY}{physical layer}
\newacronym{bler}{BLER}{block error rate}
\newacronym{tb}{TB}{transport block}
\newacronym{tbs}{TBS}{transport block size}
\newacronym{per}{PER}{packet error rate}
\newacronym{prb}{PRB}{physical resource block}
\newacronym{re}{RE}{resource element}
\newacronym{rb}{RB}{resource block}
\newacronym{sr}{SR}{scheduling request}
\newacronym{sg}{SG}{scheduling grant}
\newacronym{ack}{ACK}{acknowledgement}
\newacronym{rl}{RL}{reinforcement learning}
\newacronym{ml}{ML}{machine learning}
\newacronym{mdp}{MDP}{Markov decision process}
\newacronym{ql-amc}{QL-AMC}{Q-learning based adaptive modulation and coding}
\newacronym{ql-la}{QL-LA}{Q-learning based link adaptation}
\newacronym{dqn}{DQN}{deep Q-network}
\newacronym{ma}{MA}{multi-agent}
\newacronym{marl}{MARL}{multi-agent reinforcement learning}
\newacronym{mas}{MAS}{multi-agent systems}
\newacronym{maddpg}{MADDPG}{multi-agent deep deterministic policy gradient}
\newacronym{matd3}{MATD3}{multi-agent twin delayed deep deterministic policy gradient}
\newacronym{ddpg}{DDPG}{deep deterministic policy gradient}
\newacronym{ctde}{CTDE}{centralized training and decentralized execution}
\newacronym{sdu}{SDU}{service data unit}
\newacronym{pdu}{PDU}{protocol data unit}
\newacronym{ci}{CI}{confidence interval}
\newacronym{mlp}{MLP}{multilayer perceptron}
\newacronym{relu}{ReLU}{rectified linear unit}
\newacronym{dec-pomdp}{Dec-POMDP}{decentralized partially observable Markov decision process}
\newacronym{sc}{SC}{speaker consistency}
\newacronym{ic}{IC}{instantaneous coordination}

%% file: symbols.tex
\newglossary[nlg]{notation}{not}{ntn}{Notation}

\newglossaryentry{not:full-model}{
   name=\ensuremath{F},
   description={complete model},
   type=notation
   }

\newglossaryentry{not:decoder-model}{
   name=\ensuremath{F_{\mathrm{Dec}}},
   description={decoder model},
   type=notation
   }

\newglossaryentry{not:encoder-model}{
   name=\ensuremath{F_{\mathrm{Enc}}},
   description={encoder model},
   type=notation
   }

\newglossaryentry{not:combiner-model}{
   name=\ensuremath{F_{\mathrm{Comb}}},
   description={combiner model},
   type=notation
   }

\newglossaryentry{not:comm-system}{
   name=\ensuremath{C},
   description={communication system},
   type=notation
   }

\newglossaryentry{not:nUE}{
   name=\ensuremath{L},
   description={number of users},
   type=notation
   }

   \newglossaryentry{not:nGroups}{
   name=\ensuremath{G},
   description={number of groups},
   type=notation
   }

\newglossaryentry{not:entropy}{
      name=\ensuremath{\mathit{H}},
      description={entropy},
      type=notation
      }

\newglossaryentry{not:buffer-size}{
   name=\ensuremath{B},
   description={buffer size},
   type=notation
   }

\newglossaryentry{not:bitlength}{
   name=\ensuremath{\Upsilon},
   description={bitlength},
   type=notation
   }

\newglossaryentry{not:policy-net}{
   name=\ensuremath{\mu},
   description={policy network},
   type=notation
   }

\newglossaryentry{not:mutual-information}{
   name=\ensuremath{I},
   description={mutual information},
   type=notation
   }

\newglossaryentry{not:policy-params}{
   name=\ensuremath{\theta},
   description={policy network},
   type=notation
   }

\newglossaryentry{not:signal-overhead}{
   name=\ensuremath{\Theta},
   description={signaling overhead},
   type=notation
   }

\newglossaryentry{not:avg-packets}{
   name=\ensuremath{\lambda},
   description={average number of packets to transmit},
   type=notation
   }

\newglossaryentry{not:reward-signal}{
   name=\ensuremath{\rho},
   description={number of packets to transmit},
   type=notation
   }

\newglossaryentry{not:p-patch}{
   name=\ensuremath{p_{\mathrm{p}}},
   description={probability of having a partial observation},
   type=notation
   }

   \newglossaryentry{not:vec-query}{
   name=\ensuremath{\mu},
   description={query vector},
   type=notation
   }

   \newglossaryentry{not:generator-query}{
   name=\ensuremath{\mathcal{Q}},
   description={query generator model},
   type=notation
   }
   \newglossaryentry{not:generator-key}{
   name=\ensuremath{\mathcal{K}},
   description={key generator model},
   type=notation
   }
   
   \newglossaryentry{not:vec-key}{
   name=\ensuremath{\kappa},
   description={key vector},
   type=notation
   }

   \newglossaryentry{not:attention-weights}{
   name=\ensuremath{\mathbf{W}},
   description={attention weights},
   type=notation
   }

\newglossaryentry{not:generator-weight}{
   name=\ensuremath{\mathcal{W}},
   description={weight generator model},
   type=notation
   }

   \newglossaryentry{not:observation}{
   name=\ensuremath{o},
   description={intermediate observation},
   type=notation
   }

\newglossaryentry{not:episode-duration}{
   name=\ensuremath{T},
   description={maximum number of TTIs},
   type=notation
   }

\newglossaryentry{not:time-step}{
   name=\ensuremath{t},
   description={time step},
   type=notation
   }

\newglossaryentry{not:dl-vocabulary-size}{
   name=\ensuremath{D},
   description={downlink control message vocabulary size},
   type=notation
   }

\newglossaryentry{not:ul-vocabulary-size}{
   name=\ensuremath{U},
   description={uplink control message vocabulary size},
   type=notation
   }

\newglossaryentry{not:dl-message}{
   name=\ensuremath{m},
   description={downlink control message},
   type=notation
   }

\newglossaryentry{not:ul-message}{
   name=\ensuremath{n},
   description={uplink control message},
   type=notation
   }

\newglossaryentry{not:ue-action}{
   name=\ensuremath{a},
   description={UE action},
   type=notation
   }

\newglossaryentry{not:tbs}{
   name=\ensuremath{L_{\mathrm{TB}}},
   description={transport block size},
   type=notation
   }

\newglossaryentry{not:tti}{
  name=\ensuremath{T_{\mathrm{TTI}}},
  description={transmission time interval duration},
  type=notation
  }

\newglossaryentry{not:state}{
  name=\ensuremath{x},
  description={agent state},
  type=notation
  }

\newglossaryentry{not:history-len}{
  name=\ensuremath{K},
  description={length of history buffer},
  type=notation
  }

\newglossaryentry{not:episodes-train}{
  name=\ensuremath{N_\mathrm{train}},
  description={number of training episodes},
  type=notation
  }

\newglossaryentry{not:episodes-eval}{
  name=\ensuremath{N_\mathrm{eval}},
  description={number of evaluation episodes},
  type=notation
  }

\newglossaryentry{not:episodes-test}{
  name=\ensuremath{N_\mathrm{test}},
  description={number of test episodes},
  type=notation
  }

\newglossaryentry{not:repetitions}{
  name=\ensuremath{N_\mathrm{rep}},
  description={number of test episodes},
  type=notation
  }

\newglossaryentry{not:n-receptions}{
  name=\ensuremath{N_\mathrm{SDUs}},
  description={number of test episodes},
  type=notation
  }

\newglossaryentry{not:n-collisions}{
  name=\ensuremath{N_\mathrm{c}},
  description={number of collisions},
  type=notation
  }